\documentclass[aps,prb,twocolumn,showpacs,preprintnumbers,superscriptaddress,amsmath,amssymb,longbibliography]{revtex4-2}
\usepackage{graphicx}
\usepackage{dcolumn}
\usepackage{bm}
\usepackage{amsfonts}
\usepackage{amsmath}
\usepackage{amssymb}
\usepackage[dvipsnames]{xcolor}
\usepackage{enumerate}
\usepackage{hyperref}
\usepackage{siunitx}
\DeclareSIUnit{\fu}{{f.u.}}
\DeclareSIUnit{\euc}{{e/u.c.}}
\usepackage{color}
\usepackage[section]{placeins}
\usepackage{filecontents}
\usepackage{float}
\usepackage{subcaption}

\begin{document}

\title{Electron-vacancy scattering in SrNbO$_3$ and SrTiO$_3$: A DFT-NEGF study}

\author{Victor Rosendal}
\affiliation{Department of Energy Conversion and Storage, Technical University of Denmark, 2800 Kgs. Lyngby, Denmark}
\author{Nini Pryds}
\affiliation{Department of Energy Conversion and Storage, Technical University of Denmark, 2800 Kgs. Lyngby, Denmark}
\author{Dirch H. Petersen}
\affiliation{Department of Energy Conversion and Storage, Technical University of Denmark, 2800 Kgs. Lyngby, Denmark}
\author{Mads Brandbyge}
\affiliation{Department of Physics, Technical University of Denmark, 2800 Kgs. Lyngby, Denmark}

\date{\today}

\begin{abstract}
Oxygen vacancies are often attributed to changes in the electronic transport for perovskite oxide materials (ABO$_3$). Here, we use density functional theory (DFT) coupled with non-equilibrium Green's functions (NEGF) to systematically investigate the influence of O vacancies and also A and B-site vacancies, on the electronic transport as characterised by a scattering cross-section. We consider SrNbO$_3$ and n-type SrTiO$_3$ and contrast results for bulk and thin film (slab) geometries. By varying the electron doping in SrTiO$_3$ we get insight into how the electron-vacancy scattering vary for different experimental conditions. We observe a significant increase in the scattering cross-section (in units of square-lattice parameter, $a^2$) from ca. $0.5-2.5a^2$ per vacancy in SrNbO$_3$ and heavily doped SrTiO$_3$ to more than $9a^2$ in SrTiO$_3$ with 0.02 free carriers per unit cell. Furthermore, the scattering strength of O vacancies is enhanced in TiO$_2$ terminated surfaces by more than 6 times in lowly doped SrTiO$_3$ compared to other locations in slabs and bulk systems.
Interestingly, we also find that Sr vacancies go from being negligible scattering centers in SrNbO$_3$ and heavily doped SrTiO$_3$, to having a large scattering cross-section in weakly doped SrTiO$_3$.
We therefore conclude that the electron-vacancy scattering in these systems is sensitive to the combination of electron concentration and vacancy location.
\end{abstract}
   
\maketitle

\section{Introduction}
The plethora of material properties that perovskite oxides (ABO$_3$) offer is rewarding both for applications, but also from a fundamental point-of-view. The behavior of these materials emerges from the complex interaction among atomic, electronic, and lattice degrees of freedom
Perovskite oxide materials show many exotic properties that can be controlled by tuning the chemical composition. Through optimized growth procedures, to limit defect formation, oxide films can exhibit highly mobile electrons.~\cite{Son2010} More specifically, it has been shown both experimentally and theoretically that the oxygen content is crucial for the electronic structure and properties of oxides.~\cite{Gunkel2020,Park2020b} The disorder due to oxygen vacancies have been connected to the poor or even insulating electronic behaviour of thin film oxides.~\cite{Hu2014,Jeschke2015,Liao2015,Ha2020,Okuma2022} On the other hand, O vacancies are known to form n-type SrTiO$_3$, and they have even been attributed to the formation of highly conducting two-dimensional electron gases at the interface between insulating oxides.~\cite{Liu2013} This suggests that O vacancies both result in free electrons, but also act as scattering centers. Furthermore, both A and B-site vacancies/disorder have been correlated with electrical properties in perovskite oxides.~\cite{Liao2015, Lee2017, Wang2019} In other words, it might be important to consider the different types of vacancies and their relative importance might not be universal. Moreover, since the experimental studies are often based on thin films, the role of finite thickness should be considered. The thickness of the film could influence the vacancy formation, but perhaps also the scattering strength of one vacancy. The exact relation between vacancies, film thickness and the electron mobility still remains to be explored.

An oxide of recent interest is SrNbO$_3$.~\cite{Oka2015} The interest stems from the potential of using SrNbO$_3$ as a transparent conductor~\cite{Park2020} or even to realise extremely mobile electrons by forming semi-Dirac dispersions~\cite{Ok}. These findings make it worthwhile to ask what the effects of different vacancies are on the electronic transport properties in a material such as SrNbO$_3$. Furthermore, experiments suggest that the carrier concentration of conducting oxides is thickness dependent~\cite{Chatterjee2022, Zhu2022}, which leads to the question whether this has a significant impact on the behaviour of the electron-vacancy scattering?

To answer these questions, we examine the influence of vacancies on the electronic transport in SrNbO$_3$ and n-type doped SrTiO$_3$ using density functional theory (DFT) based calculations. We employ non-equilibrium Green's functions (NEGFs) to remove periodic boundary conditions (PBC) and to study transport. Both bulk and slabs are considered, and different vacancy types and positions are sampled. To investigate the importance of carrier concentration, we systematically vary the electron count of SrTiO$_3$ by alloying with V within the virtual crystal approximation (VCA). 

\begin{figure*}
    \includegraphics[width=\textwidth]{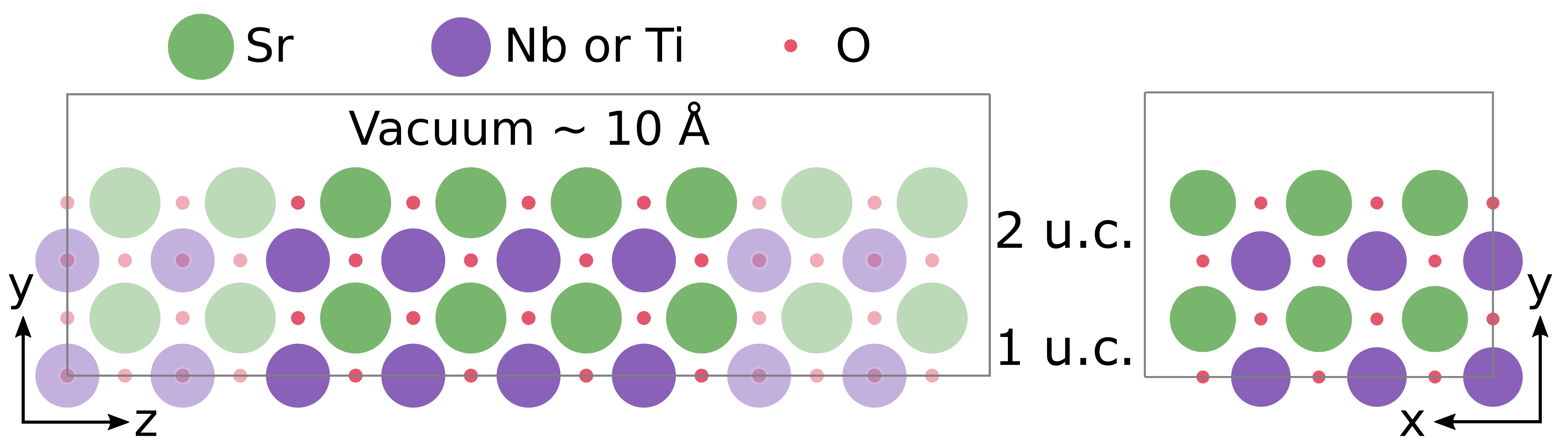}
    \caption{Schematic of the transport problem of the 2 u.c. thick slab of SrNbO$_3$ or SrTiO$_3$. The transport is in the $z$-direction and repitition of 3 u.c. is used in $x$-direction to reduce the vacancy-image interaction. The transparent circles denote the ``Left''/``Right'' electrode regions. The vacancies are placed at the center of the scattering region. Note that the vacuum region is shrunk for visualization purposes. The unit cells are counted starting from the NbO$_2$ or TiO$_2$ termination.}
    \label{fig:geometry}
\end{figure*}

We show that the vacancies reduce the electron transmission, and hence conductance, as expected. More interestingly, we note how O vacancies have a rather small scattering cross-section, and resistance, compared to Nb or Ti vacancies. Furthermore, we find that decreasing the electron concentration in SrTiO$_3$ leads to a significant increase in the scattering cross-section of Sr and Ti vacancies, while O vacancies are not affected as much. This suggests that A and B-site vacancies might play a large role in the electronic behaviour of thin film oxides, especially in conditions where the carrier concentrations are depleted in the thin films. Moreover, slabs show steps in the energy-resolved conductance due to confinement effects. Interestingly, we find that TiO$_2$ terminated surfaces show enhanced O vacancy scattering cross-sections, compared to bulk and other layers.

\section{Method}

\begin{figure*}
    \centering
    \includegraphics[width=\textwidth]{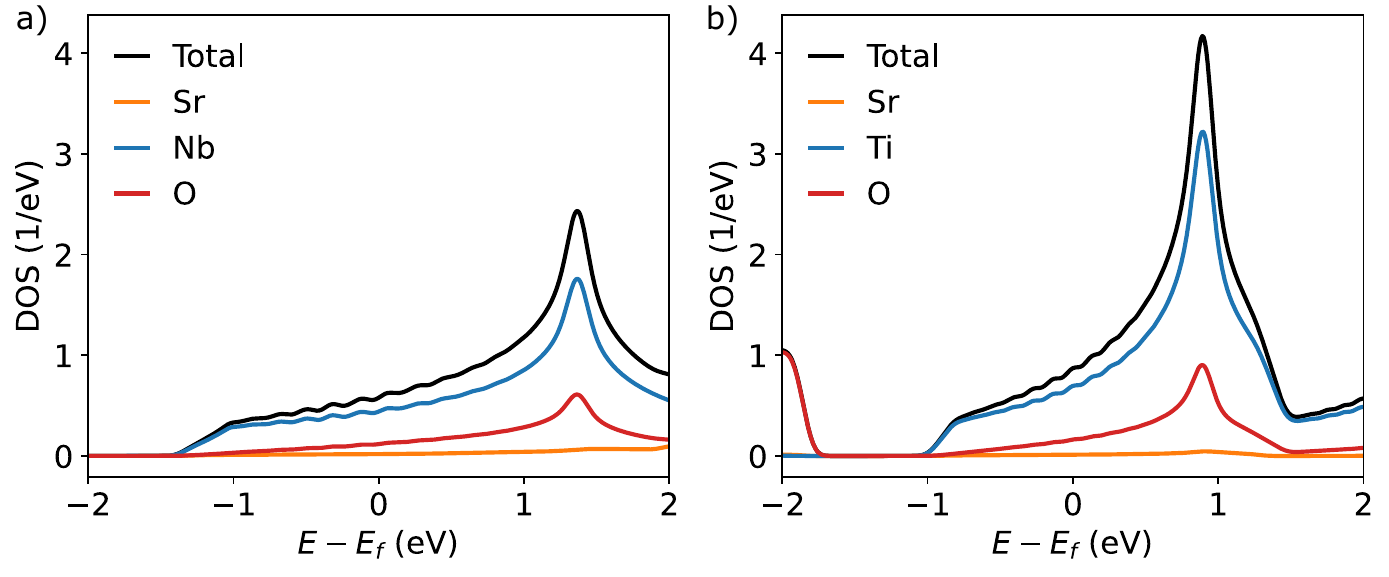}
    \caption{Atom projected DOS of bulk SrNbO$_3$ and SrTiO$_3$ doped with \SI{1}{\euc} The DOS is given for the 5 atom unit cell. Conduction bands have strong Nb/Ti and O character.}
    \label{fig:pdos}
\end{figure*}

The PBE~\cite{Perdew1996} DFT functional was used throughout the work.
To study the electron transport we use NEGF, in which one separates the ``scattering'' region (colorful circles in Figure~\ref{fig:geometry}) from the semi-infinite ``electrode'' regions (transparent circles in Figure~\ref{fig:geometry}). We consider both SrNbO$_3$ and SrTiO$_3$ doped with electrons by mixing the pseudopotential of Ti and V within the VCA.~\cite{Poloni2010}. The electrodes are pristine (bulk or slabs) cells and act as infinite reservoirs where the electrons are in (local) equilibrium described by a temperature and chemical potential. The scattering region contains repetitions of the perovskite oxide cells along the transport direction ($z$), but now with a vacancy included. This is where the periodic translational symmetry of the lattice is broken and a resistance emerges. PBC are applied in the transverse $x$ and $y$-directions (we use 3 u.c. cells in both directions in bulk and 3 u.c. in $x$-direction in slabs). Therefore, the defect-defect distance is 3 unit cells in the transverse directions. This was a compromise between the cost and accurate description of \textit{local} point defects. In the $x,y$ directions we employed 12 k-points (per unit cell) for the calculation of self-consistent charge densities, while the transmission functions were calculated using 60 k-points (per unit cell). In the NEGF technique the semi-infinite ``Left/Right'' electrodes are coupled to the central region via left/right self-energy terms.
In practice the calculations were done using Siesta~\cite{siesta} and TranSiesta~\cite{Brandbyge2002,Papior2017} which use linear combination of atomic orbitals (LCAO) basis sets which we benchmark against plane-wave calculations. A comparison between the LCAO and a plane-wave band structures can be found in Appendix~\ref{sec:appbasis}. The vacancies were introduced as ``ghost atoms'' (LCAO atoms without pseudo-potentials). Transmission functions were calculated based on the self-consistent electronic structure using TBtrans~\cite{Papior2017} and post-processing was done using SISL~\cite{Papior2023} codes. 

The electrodes were two unit cells in transport direction, which is necessary to make sure there are only matrix elements between orbitals up to nearest neighbour electrode cells. Furthermore, the NEGF formalism requires that the connection between the electrodes and the first layer in the transport direction of the central region is ``bulk-like'' and not perturbed by the defect. In other words, there should be an extension of the central region that is identical to the the electrode region where the perturbation caused by the defect can vanish. For this reason, we chose to use a central region of 4 unit cells in the transport direction, see Fig.~\ref{fig:geometry}. 
In the slab calculations, a vacuum of \SI{10}{\angstrom} was introduced to separate the slab from its images. The slabs were chosen to be asymmetric. Due to the asymmetry of the slabs we employed a dipole correction to compensate for the artificial electrostatic interaction between the two sides of the slabs.

The cubic phase of SrNbO$_3$ and SrTiO$_3$ were used with lattice parameters of $a=\SI{4.0182}{\angstrom}$ and $a=\SI{3.8958}{\angstrom}$, respectively, as predicted using DFT~\cite{Rosendal2023}. Note, that we do not consider the change in lattice parameter when electron doping SrTiO$_3$. Furthermore, we do not systematically relax the ionic degrees of freedom. This simplification is motivated by the fact that a relaxation did not affect the transmissions significantly in 2 u.c. SrNbO$_3$ without nor with O, Nb, or Sr vacancies (see Appendix~\ref{sec:relaxation}).

In this work, we limit ourselves to zero bias conditions, i.e., the linear response regime. The transmission function from left to right electrode takes the form:
\begin{equation}
    T_{LR} = \mathrm{Tr}\left[ \mathbf{G}\mathbf{\Gamma}_L \mathbf{G}^{\dagger}\mathbf{\Gamma}_R \right]
\end{equation}
where $\mathbf{G}$ is the retarded Green's function in the scattering region determined from overlap ($\mathbf{S}$) and Hamiltonian ($\mathbf{H}$) matrices in the scattering region,
\begin{equation}\label{eq:greensfunc}
    \mathbf{G} = \left[ (E + i0^+) \mathbf{S} - \mathbf{H} - \mathbf{\Sigma}_L - \mathbf{\Sigma}_R \right]^{-1}
\end{equation}
and $\mathbf{\Gamma}_{L/R}$ is the ``level broadening matrix'' due to the left or right electrodes:
\begin{equation}\label{eq:broadeningmatrix}
    \mathbf{\Gamma}_{L/R} = i\left[ \mathbf{\Sigma}_{L/R} - \mathbf{\Sigma}_{L/R}^{\dagger} \right] 
\end{equation}
We have omitted the energy and momentum variables for simplicity. In Eqs.~\ref{eq:greensfunc} and \ref{eq:broadeningmatrix}, the electrode self-energy is denoted $\mathbf{\Sigma}_{L/R}$. We present our results using the conductance function, $G(E) = \frac{e^2}{h} T_{LR}(E)$, normalised with the cross-sectional ($xy$) area of the conductor, $A$, in units of primitive unit cell area ($a^2$).
The conductance is given per spin channel, there are therefore two identical channels with equal conductances since we perform unpolarized calculations. We note that we have observed a small spin selective electron-vacancy scattering. However, while this could be an interesting direction for future studies, we neglect this effect due to the computational complexity and observed sensitivity to ionic relaxation.

\section{Electronic structure of $\mathrm{SrNbO}_3$ and $\mathrm{SrTiO}_3$}\label{sec:elecstruct}
Charge is conducted by electrons near the Fermi level ($E_F$) and it is therefore insightful to examine the electronic structure in the vicinity of the Fermi level. To do this, we calculated the atomic projected density of states (PDOS) of bulk SrNbO$_3$ and SrTiO$_3$ doped with \SI{1}{\euc} The PDOS is shown in Figure~\ref{fig:pdos}. Note, the results correspond to the cubic 5 atom unit cells without any vacancies. It is therefore these states that will conduct charge in the pristine cases. The conduction bands are qualitatively similar, however, the bandwidth is substantially larger for SrNbO$_3$ compared to SrTiO$_3$. This is due to the more delocalized character of the Nb $4d$ states compared to the smaller Ti $3d$ states. To a large extent, the conduction bands originate from the atomic orbitals of the B-site ions (Nb and Ti). There is also a substantial contribution from the O orbitals, while the contribution of Sr on the conduction bands is negligible. From the PDOS, we expect the materials to be conducting, and that the conduction will be done by the B-site $d$ states that are bridged by the O ($2p$) states. The atomic origins of the states (naively) suggest that B-site and O vacancies should dominate the scattering.

The PDOS in Figure~\ref{fig:pdos} were calculated using LCAO DFT, which agree very well with previous plane-wave DFT, see Appendix~\ref{sec:appbasis} and Reference~\cite{Rosendal2023}. Furthermore, recent comparisons between plane-wave DFT and angle-resolved photoemission spectroscopy show good agreement.~\cite{Bigi2020,Chikina2023} This give us confidence in the LCAO description used here. We note that in both aforementioned studies, hole doping was likely observed, likely due to vacancies.

\section{Electron transport in $\mathrm{SrNbO}_3$}
We begin by examining the electron transport in bulk and slabs of SrNbO$_3$. We will use this as the baseline for conducting $d^1$ perovskite oxides, i.e., systems with very high electron concentration and good electrostatic screening. The influence of carrier concentration will be considered in the subsequent section.

\begin{figure*}%
    \centering
    \includegraphics[width=\textwidth]{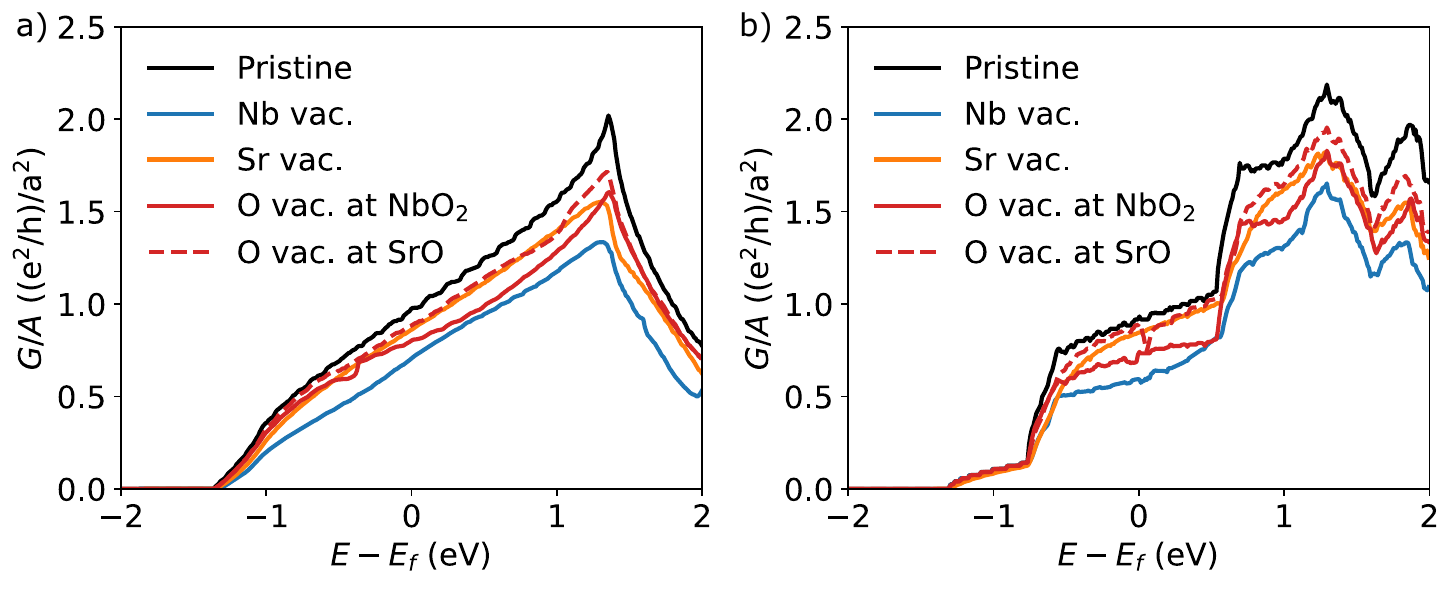}
    \caption{Conductance per unit cell area for different vacancy configurations. (a) Bulk SrNbO$_3$ with periodic boundary conditions in the transverse ($xy$) directions and a 3 u.c. supercell. (b) Two unit cell thick slab configuration of SrNbO$_3$. The vacancies are placed in the first u.c. counting from the NbO$_2$ termination layer. Periodicity is applied in the transverse ($x$) direction using a 3 u.c. supercell.}%
    \label{fig:SNO-bulk-and-slab}%
\end{figure*}

\subsection{Bulk conductance in $\mathrm{SrNbO}_3$}
Figure~\ref{fig:SNO-bulk-and-slab}a) shows the area specific electrical conductances for bulk SrNbO$_3$ with different vacancies. While charge is conducted mainly by electrons very close to the Fermi level ($E_F$), we show a larger range of energies. As expected, there is good conduction in pristine SrNbO$_3$ near the Fermi level due to the wide conduction bands originating from hybridized Nb $4d$ and O $2p$ states. Introducing vacancies reduces the conductance per area around $E_F$ by 10-30\%, as shown Figure~\ref{fig:SNO-bulk-and-slab}a). In other words, a vacancy causes resistance. Furthermore, we observe that introducing a vacancy in a NbO$_2$ layer (i.e. Nb vacancy or O vacancy between Nb ions in the transport direction), causes more resistance than a vacancy in a SrO layer (i.e. Sr vacancy or O vacancy between Nb ions perpendicular to the transport direction). This is expected due to the origin of the conduction states in SrNbO$_3$ mentioned earlier, i.e., the conduction in done by Nb states ``connected'' by O states.

\begin{figure}
\includegraphics[width=0.5\textwidth]{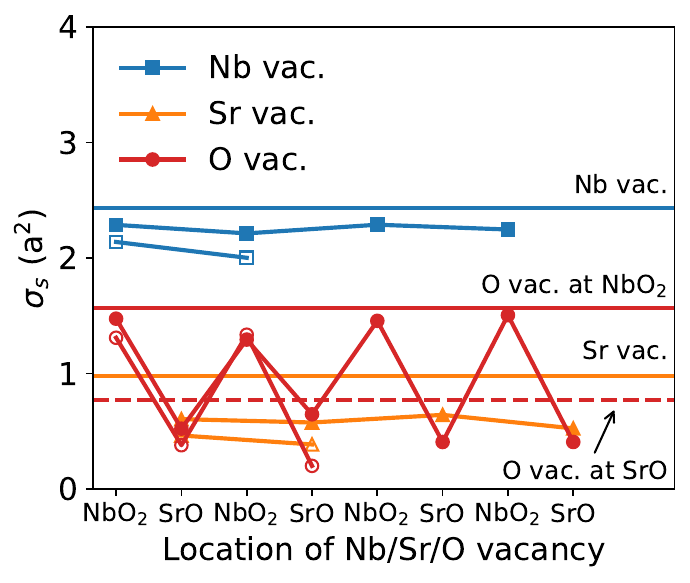}
    \caption{Scattering cross-sections of different vacancies in SrNbO$_3$. The values were evaluated at the Fermi level, and the cross-sections are given in units of the unit cell area. Horizontal lines correspond to bulk values and the hollow (filled) markers correspond to scattering from different vacancy locations in 2 (4) u.c. thick slabs.
    }
    \label{fig:SNO-scattering}
\end{figure}

\subsection{Slab conductance in $\mathrm{SrNbO}_3$}
Next, we investigate the impact of confinement and surface (present in slabs) on the conductive properties of SrNbO$_3$. Asymmetric slabs of 2, 3, and 4 u.c. SrNbO$_3$ were investigated. In Fig.~\ref{fig:SNO-bulk-and-slab}b), the conductances of a 2 u.c. slab of SrNbO$_3$ with different vacancies are shown. The confinement leads to large jumps in the conductance depending on the slab thickness (see Appendix~\ref{sec:confinement} for 2, 3 and 4 u.c. pristine SrNbO$_3$) due to the quantization transverse to the slab plane.
Here, we consider vacancies placed within the first unit cell on the NbO$_2$ terminated side, see Figure~\ref{fig:geometry}. Other vacancy locations will be discussed later.
We find that the trend in scattering seen as drop in conductance is quite similar to the bulk cases; vacancies in a SrO layer are much less disruptive for electron transport than those in an NbO$_2$ layer. Qualitatively, vacancies placed in the 2nd u.c. from the NbO$_2$ termination (see Fig.~\ref{fig:geometry}) show similar trends in resistances, or drops in conductances. For quantitative comparisons, we refer to the following section.

\subsection{Vacancy scattering cross-sections in SrNbO$_3$}
Having shown the energy resolved conductance for bulk and one slab configuration of SrNbO$_3$, we now extract the scattering cross-sections from vacancies placed in bulk and slabs of different thicknesses at different locations. We define the scattering cross-section as,
\begin{equation}
    \sigma_s = A(T_0(E_F) - T_d(E_F))/T_0(E_F)
    \label{eq.sigma}
\end{equation}
where $A$ is the geometrical cross-section, and $T_0$ and $T_d$ are the area-specific transmissions of the pristine and defective system, respectively.~\cite{Markussen2007, Markussen2010} The scattering cross-section gives an intuition of how much of the cross-sectional area is effectively lost due to the introduction of the vacancies. This is done for a certain transverse supercell, hence its maximum value will be $A$ corresponding to a blocked, non-transmitting supercell. With this definition of cross-section we will get a lower estimate. Note that due to the finite sampling of the transmission functions there might be numerical artifacts which could propagate into the evaluated scattering cross-sections. Therefore, we smear the transmissions with the energy derivative of the Fermi-Dirac distribution at \SI{300}{\kelvin}. This is only done when evaluating scattering cross-sections and subsequent quantities (e.g., mobility later on).
 
Figure~\ref{fig:SNO-scattering} presents the scattering cross-sections for 2 and 4 u.c. SrNbO$_3$. For simplicity, the results of 3 u.c. are omitted. Different transverse locations in the slab (i.e., different unit cells in $y$-direction in Fig.~\ref{fig:geometry}) and vacancy types are sampled. Furthermore, the bulk values are shown as horizontal lines. As noted above the trend is similar for bulk and slab configurations and there is no strong layer dependency on the scattering strengths, except for the O vacancy positions where the oscillating behaviour of the O scattering cross-section depicts the difference in scattering strength from vacancies in NbO$_2$ and SrO layers. 
Overall, it is clear that in SrNbO$_3$ the Nb vacancies have the biggest impact on the transport, followed by O vacancies in NbO$_2$ layers and finally Sr and O vacancies in SrO layers have negligible impact. Bulk scattering cross-sections are ca. $2.5a^2$ for Nb vacancies, $1.6a^2$ for O vacancies in NbO$_2$ layer, $1.0a^2$ for Sr vacancies, and ca. $0.8a^2$ for O vacancies in SrO layers. For the ultra thin slabs the scattering cross-sections are ca. $2.1a^2$ for Nb vacancies, $1.4a^2$ for O vacancies in NbO$_2$ layer, $0.5a^2$ for Sr vacancies, and ca. $0.5a^2$ for O vacancies in SrO layers.
It is interesting to note that the scattering cross-sections are generally lower for the slab compared to bulk. This may be due to the discretization in terms of transverse quantization which limits the scattering channels.

\section{Electron transport in n-type $\mathrm{SrTiO}_3$}
Next we will contrast the results from the metallic SrNbO$_3$ to semi-conducting SrTiO$_3$, to explore whether the concentration of electrons affects the electron-vacancy scattering in n-type SrTiO$_3$. Intuitively, the total conductance should increase with more electrons. This is exactly what we observe when doping SrTiO$_3$ with vanadium (V), see Figure~\ref{fig:STO-bulk} in Appendix~\ref{sec:doping}. We here dope with V (instead of e.g. Nb), since V has $3d$ orbitals like Ti. A systematic shift in the Fermi level is observed when doping with V, and the conduction band gets occupied.
In this section, we focus on examining the influence of electron concentration on the scattering mechanism, rather than on the more straightforward relationship between the number of free carriers and the conductance.

\begin{figure*}
    \centering
    \includegraphics[width=\textwidth]{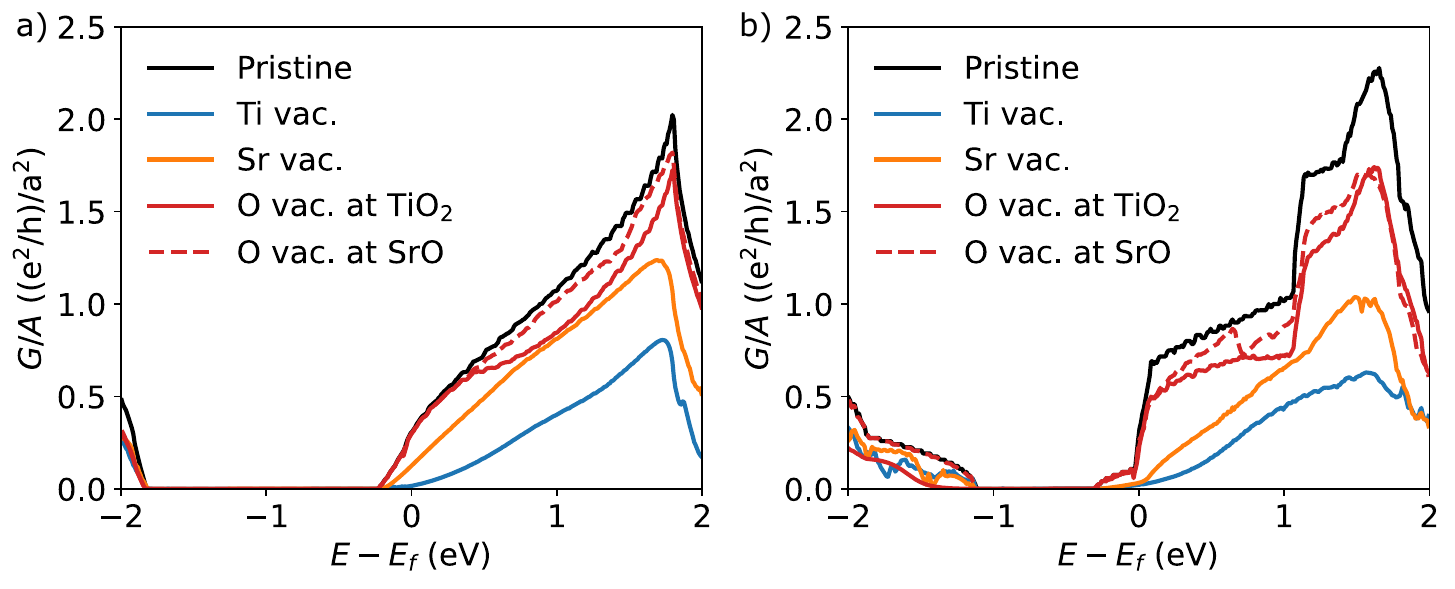}
    \caption{Conductance per unit cell area for different vacancy configurations. (a) Bulk SrTiO$_3$ doped with \SI{0.1}{\euc} with periodic boundary conditions in the transverse ($xy$) directions and a 3 u.c. supercell.  (b) Two unit cell thick slab configuration of SrTiO$_3$ doped with \SI{0.1}{\euc}. The vacancies are placed in the first unit cell counting from the TiO$_2$ termination layer. Periodicity is applied in the transverse ($x$) direction and a 3 u.c. supercell.}%
    \label{fig:STO-bulk-and-slab}%
\end{figure*}

\subsection{Bulk conductance in doped SrTiO$_3$}
To illustrate the conductive properties of n-type bulk SrTiO$_3$, we present the conductance of SrTiO$_3$ with \SI{0.1}{\euc} in Figure~\ref{fig:STO-bulk-and-slab}a). The pristine system resembles SrNbO$_3$, however with a shift in Fermi level and a smaller conduction band width. The narrower bands of SrTiO$_3$ are expected due to the more localized nature of the Ti $3d$ states compared to the Nb $4d$ states. Interestingly, in comparison to SrNbO$_3$, the decrease in conductance caused by a Sr vacancy in SrTiO$_3$ is significantly higher (by ca. \SI{200}{\percent} for \SI{0.1}{\euc}), whereas the scattering resulting from an O vacancy is even less pronounced. Similar to the observations in SrNbO$_3$, the vacancy at the B-site continues to have the most substantial effect on the conductance.

\subsection{Slab conductance in doped SrTiO$_3$}
Next, the transport in 2 u.c. thick slabs of SrTiO$_3$ doped with \SI{0.1}{\euc} is investigated. Like in the case of SrNbO$_3$, we only present the transport in slabs with vacancies placed in or near the TiO$_2$ termination layer (i.e. vacancies in the first unit cell in the $y$-direction cf. Fig.~\ref{fig:geometry}). The conductances are presented in Figure~\ref{fig:STO-bulk-and-slab}b). The comparison between the bulk SrTiO$_3$ and the 2 u.c. thick slab of SrTiO$_3$ shows a similarity to the comparison observed between bulk and slab SrNbO$_3$. In both cases, confinement leads to distinct steps in conductance, but the overall trends regarding the impact of different types of vacancies remain consistent. 
The somewhat remarkable finding that the O vacancies now play a minor role can be understood from the Fermi level position being at the band edge, rather than well inside the band as in SrNbO$_3$. At higher energies the O vacancy scattering comes into play.
In essence, there appears to be no qualitative difference in electron-vacancy scattering when comparing bulk and slab forms of SrTiO$_3$, except we note that the on-set of O vacancy scattering is much earlier with respect to energy in the slab compared to bulk.

\begin{figure*}
    \includegraphics[width=\textwidth]{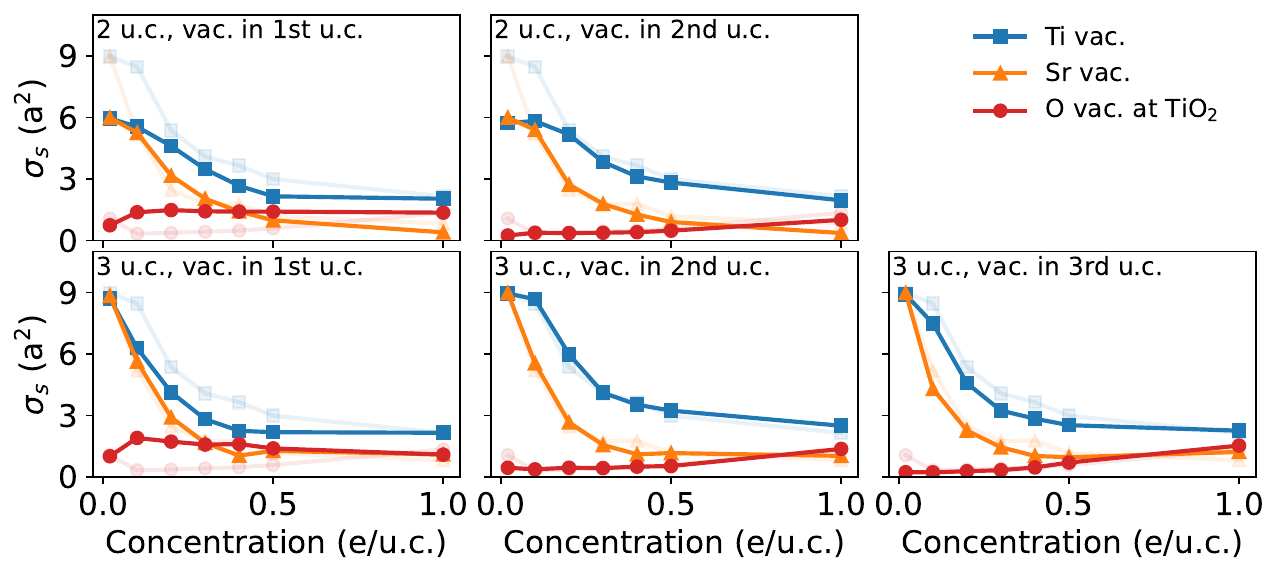}
    \caption{Scattering cross-section due to vacancies in doped SrTiO$_3$ with varying number of extra electrons. The two rows correspond to 2 u.c. and 3 u.c. thick slabs, respectively. Bulk values are included as transparent lines. The columns correspond to the different layers in which the vacancies are placed in (from TiO$_2$ termination to the left to the SrO termination to the right). Note, that the geometrical cross-section of the 2 u.c. slab differs from the others ($A = \text{width} \times \text{height}$).}
    \label{fig:STO-thickness-conc-scatter}
\end{figure*}

\subsection{Doping and scattering cross-sections in SrTiO$_3$}
To quantitatively investigate the relation between scattering cross-section and carrier concentration, we systematically evaluated the conductance of SrTiO$_3$ with varying doping and defect configurations. In Figure~\ref{fig:STO-thickness-conc-scatter}, the scattering cross-sections are shown versus carrier concentration for bulk, 2 u.c. and 3 u.c. thick SrTiO$_3$, respectively. The layered resolved scattering cross-sections are shown in the slab cases. We first summarize the finding from the bulk results (transparent lines in Fig.~\ref{fig:STO-thickness-conc-scatter}), then move over to nuances in the slab results. While we here focus on scattering at Fermi level, energy resolved conductances versus doping for the defective bulk systems can be found in Appendix~\ref{sec:doping}. In highly doped bulk (\SI{1.0}{\euc}), the order and magnitudes of the scattering cross-sections are similar to those of SrNbO$_3$. Specifically, Ti vacancies result in the largest resistance and scattering cross-section, followed by O vacancies in the TiO$_2$ layer, and finally Sr vacancies. Interestingly, reducing the number of electrons causes the scattering cross-section to increase for Ti and Sr vacancies. The scattering by O vacancies is initially reduced with a decrease in the number of electrons followed by an increase between ca. \SI{0.10}{\euc} and \SI{0.02}{\euc} We note that the bottom of the conduction band can be more sensitive to finite sampling effects, and while we smear the transmission functions to evaluate the cross-sections, we do expect larger inaccuracies in the lower doped systems. Between \SI{0.10}{\euc} and \SI{0.50}{\euc} the scattering cross-section is below $0.65a^2$. It is noteworthy that the scattering cross-sections of Ti and Sr vacancies increase significantly from $2.2a^2$ and $0.7a^2$, respectively, at \SI{1.0}{\euc} up to $9a^2$ at a lower doping of \SI{0.02}{\euc} Note, that the bulk geometrical cross-section is $A = 3a\times3a = 9a^2$, i.e., a scattering cross-section of $9a^2$ indicates that the transport through the defective system is zero at this defect density, and that $\sigma_s$ will be beyond $9a^2$. This is observed for the lowly doped bulk SrTiO$_3$ with Sr and Ti vacancies. 

Numerous similarities are observed when we compare the outcome of the 2 u.c. doped SrTiO$_3$ with its bulk counterparts. For instance, there is an increase in scattering due to Sr and Ti vacancies with a decrease in carrier concentration. Note, that in 2 u.c. the geometrical cross-section is $A = wh = 3a \times 2a = 6a^2$, where the $w$ and $h$ are the widths and heights of the computational cell, respectively. Hence, the ``truncated'' values ($6a^2$) indicate defect configurations with vanishing transmission. Interestingly, the O vacancy scattering is enhanced if the vacancy is placed in the TiO$_2$ termination layer compared to the TiO$_2$ layer near the SrO termination layer (noted as ``2 u.c., vac. in 1st u.c.'' and  ``2 u.c., vac. in 2nd u.c.'', respectively, in Fig.~\ref{fig:STO-thickness-conc-scatter}).
Furthermore, it seems that the Ti vacancy scattering is suppressed in the TiO$_2$ termination compared to bulk and other layers. These trends are also present in the 3 u.c. and 4 u.c. (not shown here) thick slabs. This indicates that the oxygen states near the TiO$_2$ termination have a greater impact on electronic transport, compared to oxygen states in bulk and near SrO termination layers. Hence, there is a shift in importance from Ti vacancies to O vacancies when comparing the TiO$_2$ termination layers with bulk and other layers. This was not observed in SrNbO$_3$ which indicates that the electron-vacancy is a combination of location \textit{and} carrier concentration.

From the 4 u.c. thick slabs of doped SrTiO$_3$, we observe similar trends in the scattering found in the 2 and 3 u.c. thick slabs. For SrTiO$_3$ with \SI{0.02}{\euc}, the scattering cross-sections of the Sr and Ti vacancies placed in the first unit cell (counting from TiO$_2$ termination) are $10.3a^2$ and $10.2a^2$, respectively. These values are smaller than the geometrical cross-sectional area of $A = wh = 3a \times 4a = 12a^2$, which indicates that the scattering cross-sections are in fact below $12a^2$ for these configurations. The other vacancy locations still exhibit values of $12a^2$ for SrTiO$_3$ with \SI{0.02}{\euc}, which indicates that the cross-sections are larger than $12a^2$ for Sr and Ti vacancies placed in layers further away from the TiO$_2$ termination unit cell.

These findings indicate that while TiO$_2$ termination layers seem more sensitive to O vacancies, the majority of vacancy configurations in slabs are quite similar to the bulk counterparts. More generally, the findings from this section suggests that to a large extent it is the electron concentration, and not confinement effects, that determines the impact of the electron-vacancy scattering in these oxides. Varying the carrier concentration shifts the relative importance of the different vacancies.
Furthermore, the shift in relative importance of the different vacancies can have two origins: either the atomic orbitals involved in the transport is changed with Fermi level shift or the electrostatic screening of the vacancy potential is sensitive to the number of electrons. To understand the origin, we returned to the atomistic PDOS of bulk SrTiO$_3$ (see Fig.~\ref{fig:pdos}) and examined the energy resolved conductances of bulk SrTiO$_3$ with various vacancies (see Appendix~\ref{sec:doping}). As discussed in Section~\ref{sec:elecstruct}, the conduction bands of SrTiO$_3$ have a strong Ti character, followed by O and a negligible Sr character (see Fig.~\ref{fig:pdos}b)). While Ti ions have large effect on transport \textit{and} large contribution to conduction states, Sr vacancies are the opposite. Sr vacancies have large impact on transport, however, the Sr PDOS is negligible in the conduction bands. This is highlighted in Figure~\ref{fig:STO-bulk-Sr-vac}, where the energy resolved conductance of bulk SrTiO$_3$ with Sr vacancy is shown for varying doping. The conductance of the defective system is drastically reduced when the electron doping is reduced, while the Sr PDOS remains negligible throughout the conduction band (see Fig.~\ref{fig:pdos}). This indicates that the change in scattering cross-section is not simply due to a change in which states are involved in the transport, but also how the electrons interact with the defect potential, especially in the case of Sr vacancies.

\section{Vacancies and oxide thin films}
\begin{figure}
    \centering
    \includegraphics[width=0.5\textwidth]{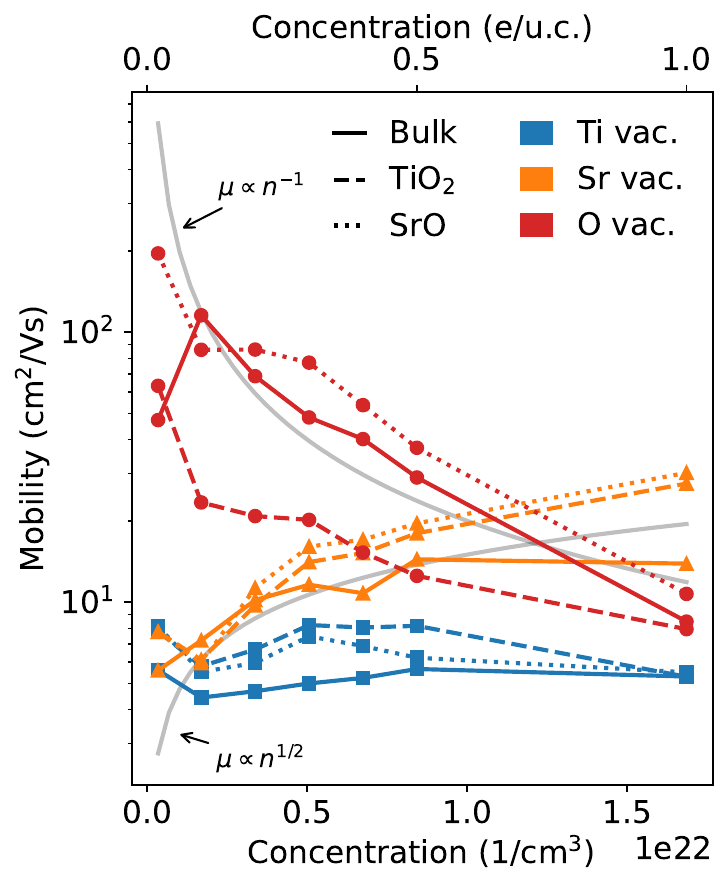}
    \caption{Vacancy limited electron mobility of SrTiO$_3$ as a function of carrier concentration. The vacancy concentrations were set to \SI{1e21}{\per\cubic\centi\metre}. The solid lines correspond to bulk results, while dashed/dotted lines correspond to 2 u.c. results with vacancies placed in unit cells closest to TiO$_2$/SrO termination layer. The gray lines are included as guides for the eye.}
    \label{fig:STO-mobility}
\end{figure}

This theoretical study was inspired by the frequent references to vacancies, particularly oxygen vacancies, as crucial factors in explaining the electrical behavior of thin film oxides. On one side, vacancies can serve as dopants (for example, oxygen vacancies in SrTiO$_3$ make it n-type), but on the other, they disrupt the periodic lattice, leading to electron scattering. The literature has yet to address how strongly a single vacancy scatters electrons in conducting perovskite oxides and, importantly, what factors affect the strength of this scattering. 
Experimental trends often show a reduction of mobility and conductivity with respect to decreased thickness. Interestingly, we observe that the TiO$_2$ terminated surfaces of lowly doped SrTiO$_3$ are more sensitive to O vacancies (more than 6 times larger scattering cross-section) compared to other layers and bulk counterparts. This is accompanied by a reduction of the scattering cross-section of the Ti vacancies in the TiO$_2$ terminated surfaces. This indicates that the electron-vacancy scattering in TiO$_2$ terminated surfaces of SrTiO$_3$ might be unique, and that O vacancies might become stronger scattering centers in these surfaces. This was not observed in SrNbO$_3$ (and highly doped SrTiO$_3$), which suggests that the electron-vacancy scattering is sensitive to an interplay between the number of free electrons and to the surface effects.
Moreover, the vacancy formation can be substrate-driven (e.g. via strain~\cite{Aschauer2013}), which could make the ultra thin films more prone to vacancies and, therefore, result in a higher total electron scattering and reduced mobility.

We have showed that the electron concentration in SrTiO$_3$ plays a significant role in determining the strength of electron-vacancy scattering, particularly for Sr and Ti vacancies. Given that experimentally measured carrier concentrations frequently vary with film thickness, it leads us to speculate that thin film oxides might exhibit increased resistivity due to a decrease in free electrons, but importantly perhaps also due to an increased electron-vacancy scattering strength due to the decreased carrier concentration.
In other words, reducing the carrier concentration could be two-fold: mainly by directly reducing the number of free carriers, and secondly by increasing the A and B-site electron-vacancy scattering strength. While O vacancies are often depicted as mobility-limiting scattering centers, we surprisingly find that the scattering cross-section of a O vacancy is often smaller than both A and B-site vacancies (except for SrNbO$_3$ and highly doped SrTiO$_3$ where all vacancies have relatively small cross-sections). This could be part of the reason why the electron gas at, e.g., the LaAlO$_3$/SrTiO$_3$ interface has such a high mobility~\cite{Ohtomo2004} even though O vacancies play a crucial role in the formation of the electron gas~\cite{Liu2013}. Perhaps the gain in extra electrons due to an oxygen vacancy outweighs the additional electron scattering thanks to the small scattering cross-section of the oxygen vacancy?

We want to point out that in this paper we have only considered single vacancies, i.e., the resistances and scattering cross-sections are due to \textit{one} vacancy in the transport direction. In reality, there will likely be a distribution of vacancies between the source and drain that lead to a total electrical resistance or mobility. In the following, we estimate the vacancy limited electron mobility for different vacancy and doping conditions. This is done to contextualize our findings and to ease the comparison between our predictions and experimental data.

\subsection{Vacancy limited mobility}

A simple estimate of a mean free path (MFP), $\ell$, based on independent defects can be found from the scattering cross-section $\sigma_s(i)$ and impurity concentration, $N_i$, of the different vacancy species $i$, \cite{Markussen2010,Saloriutta12} 
\begin{equation}
  \ell\approx 1/\sum_i(N_i\sigma_s(i)).
\end{equation}
Using this and the pristine transmission $T_0(E_F)$ one may further obtain an estimate for the conductivity of the metallic systems,
\begin{equation}
    \sigma \approx \frac{2e^2}{h} \frac{T_0(E_F)}{A}\,\ell.
\end{equation}
From the conductivity, the vacancy limited mobility can be estimated as $\mu = \sigma/ ne$. Figure~\ref{fig:STO-mobility} presents the vacancy limited mobility of SrTiO$_3$ as a function of electron concentration. The MFPs were evaluated with vacancy concentrations $N_i = \SI{1e21}{\per\cubic\centi\metre}$ for each vacancy type. Only one vacancy type was consider at a time. Note that $\mu \propto 1/N_i$, i.e., from Figure~\ref{fig:STO-mobility} the mobility can be scaled to represent other vacancy concentrations depending on experimental conditions (typical oxygen vacancy concentrations are between \SI{1e17}{\per\cubic\centi\metre}-\SI{1e22}{\per\cubic\centi\metre}~\cite{Gunkel2020}). For the given vacancy concentrations the predicted vacancy limited mobilities are between \SI{4}{\centi\metre\squared/\volt\second} and \SI{200}{\centi\metre\squared/\volt\second}. This range is comparable with experimental values for SrNbO$_3$~\cite{Oka2015} and SrTiO$_3$ in the high doping limit~\cite{Trier2018}. This indicates that real films typically have a vacancy concentration comparable to \SI{1e21}{\per\cubic\centi\metre}. Note, this is excluding any other types of defects and electron scattering mechanisms, i.e., the vacancy concentration is an overestimation. From Figure~\ref{fig:STO-mobility}, it is seen that the Ti vacancies limit the electron mobility the most (as expected by the large scattering cross-sections). The Ti vacancy limited mobility remains almost constant with respect to electron concentration. Sr vacancies have a slightly smaller impact on mobility, ca. \SI{200}{\percent} higher mobility than Ti vacancy, except for in lowly doped systems. The Sr vacancies go from being a small limitation on the mobility in highly doped systems, to being very important in lowly doped systems. This change is ca. \SI{200}{\percent} going from a carrier concentration of \SI{3.4e20}{\per\cubic\centi\metre} to \SI{1.6e22}{\per\cubic\centi\metre}, which indicates the sensitivity of electron concentration on the role of Sr vacancies. Furthermore, the role of O vacancies is also sensitive to electron concentration. The O vacancy limited mobility decreases by an order of magnitude, when the carrier concentration increases from \SI{3.4e20}{\per\cubic\centi\metre} to \SI{1.6e22}{\per\cubic\centi\metre}. While the Ti and Sr vacancy limited mobilities of 2 u.c. slabs are comparable to the bulk counterparts, the O vacancy placed in TiO$_2$ termination layer sticks out (red dashed line in Fig.~\ref{fig:STO-mobility}). The mobility can be up to 6 times lower when the O vacancy is located in a TiO$_2$ termination layer compared to bulk. Furthermore, the SrO termination layer is less sensitive to O vacancies compared to bulk. This indicates that TiO$_2$ terminated SrTiO$_3$ surfaces might be more prone to reduced mobility due to O vacancies compared to bulk and SrO terminated SrTiO$_3$. Finally, we note that since the mobility with respect to electron concentration is different for the anion and cation vacancies (as highlighted by the grey lines in Fig.~\ref{fig:STO-mobility}), mobility vs. concentration trends could potentially be used to understand which vacancies are present and also what the vacancy concentrations are.

\section{Outlook}
Transport calculations based on DFT has here been used to study the role of different vacancies on the electronic transport in oxides. While we have been able to estimate the scattering strengths and electron mobilities due to vacancies, we have made various simplifications along the way. In no particular order, it could be insightful to investigate the effect of electron correlations, e.g., by Hubbard U corrections, or to include temperature effects by considering electron-phonon scattering. Furthermore, including spin degree of freedom could be relevant for understanding how magnetic moments are stabilized -- and perhaps how to generate spin-currents via different scattering mechanisms. These are topics for future studies that hopefully will further enrich the understanding of electron transport in oxides.

\section{Conclusion}
The effect of vacancies on the electronic transport in bulk and ultra thin slabs of SrNbO$_3$ and n-type SrTiO$_3$ has been investigated. The transport problem was investigated numerically using DFT-NEGF. We find that pristine bulk and slabs are conductive and the inclusion of vacancies reduces the conductance, as expected. Furthermore, we see that while the order in terms of scattering strength among the different defects is roughly preserved, the strength in SrTiO$_3$ slabs -- especially of O vacancies depends strongly on position and doping. In particular, it can in some cases be significantly higher than the bulk estimate. As an example, O vacancy placed in TiO$_2$ termination layers have scattering cross-sections ca. 6 times larger than bulk counterparts in lowly doped SrTiO$_3$.
Moreover, we observe a strong increase in the scattering cross-section of both A and B-site vacancies when reducing the electron concentration in n-type SrTiO$_3$. These findings could be part of the explanation for the poor electrical performance of some thin film oxides, especially in conditions where the carrier concentration is depleted (which is often observed). Surprisingly, while the scattering cross-sections of O vacancies are comparable to the values of A and B-site vacancies in SrNbO$_3$ and heavily doped SrTiO$_3$, their relative scattering strengths reduce substantially in lowly doped SrTiO$_3$. This could render O vacancies rather unimportant scattering centers in some oxides, especially since they also contribute with additional free electrons which enhances the total conductance. By examining PDOS, we speculate that the change in relative importance of the vacancies are related to a change in the screening with doping and not due to a change in orbital character of the conduction states.

\appendix
\section{Basis set benchmark}\label{sec:appbasis}
\begin{figure*}
    \centering
    \includegraphics[width=\textwidth]{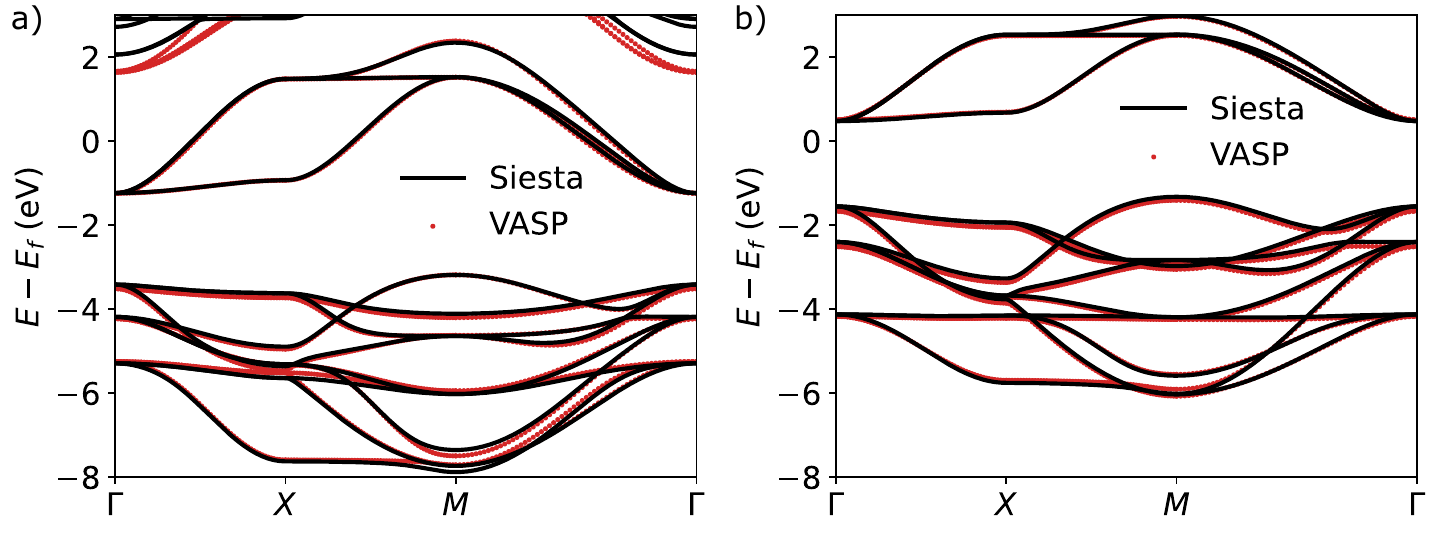}%
    \caption{Comparison between band structure of cubic (a) SrNbO$_3$ and (b) SrTiO$_3$ as calculated using Siesta and VASP. The VASP calculations were performed as described in Reference~\cite{Rosendal2023}. Note, the valence bands from VASP where shifted to fit the Siesta bands.}%
    \label{fig:siestavsvasp}%
\end{figure*}

To generate a good Siesta LCAO basis set, we benchmarked our band structures against a VASP plane-wave solution. The VASP calculation was done as described in Reference~\cite{Rosendal2023}. The resulting band structures are shown in Figure~\ref{fig:siestavsvasp}. Excellent agreement is observed. To get good agreement between the two solutions, it was necessary to use triple-zeta polarized basis functions for O. For Nb/Ti and Sr double-zeta polarized was sufficient.

\begin{figure}
    \centering
    \includegraphics[width=0.5\textwidth]{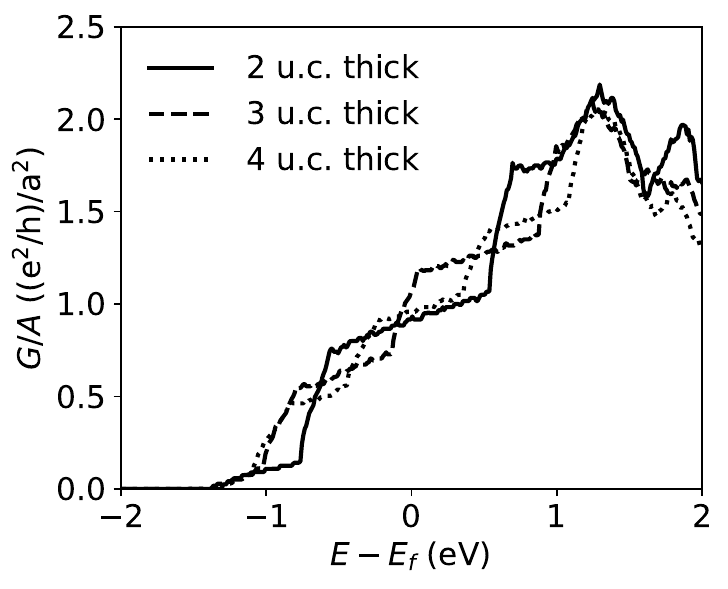}
    \caption{Conductance of pristine SrNbO$_3$ slabs with varying thickness.}
    \label{fig:SNO-thickness}
\end{figure}

\section{Conductance and confinement effects}\label{sec:confinement}
Influence of finite thickness on the electronic conductances in SrNbO$_3$ is presented in Figure~\ref{fig:SNO-thickness}. The out-of-plane confinement of the electron wavefunction leads to sharp steps in the conductances. The finite slopes remain due to the PBC in-plane (i.e., in $x$-direction in Figure~\ref{fig:geometry})

\begin{figure}
    \centering
    \includegraphics[width=0.5\textwidth]{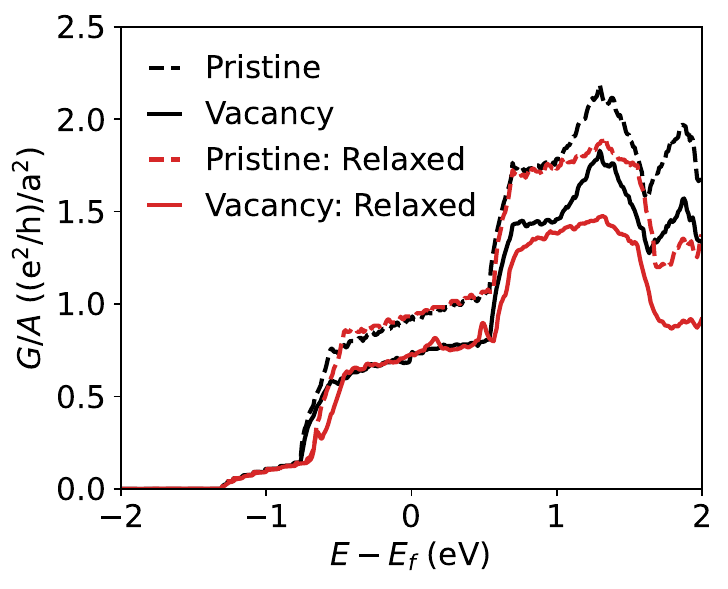}
    \caption{Conductance of 2 u.c. SrNbO$_3$ slabs without and with ionic relaxation. Here an O vacancy is placed in the NbO$_2$ termination layer.}
    \label{fig:SNO-relaxO}
\end{figure}

\section{Effect of relaxation on transport}\label{sec:relaxation}
\begin{figure}
    \centering
    \includegraphics[width=0.5\textwidth]{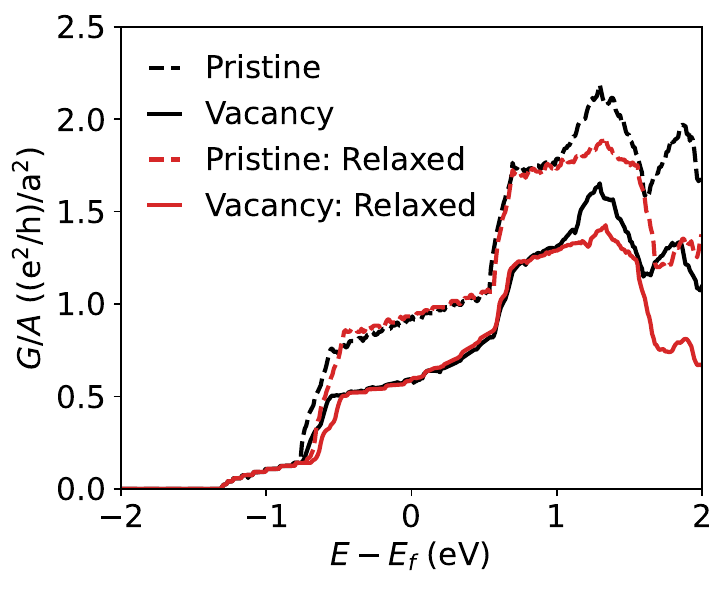}
    \caption{Conductance of 2 u.c. SrNbO$_3$ slabs without and with ionic relaxation. Here an Nb vacancy is placed in the NbO$_2$ termination layer.}
    \label{fig:SNO-relaxNb}
\end{figure}

\begin{figure}
    \centering
    \includegraphics[width=0.5\textwidth]{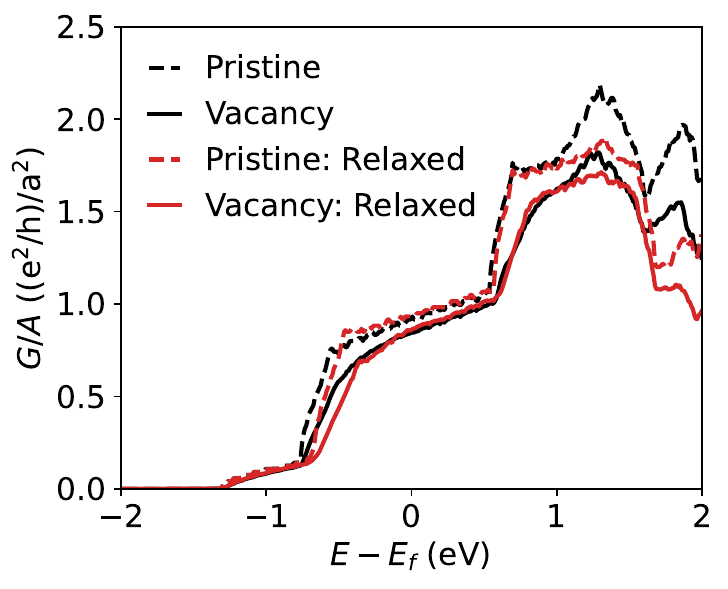}
    \caption{Conductance of 2 u.c. SrNbO$_3$ slabs without and with ionic relaxation. Here an Sr vacancy is placed closest to the NbO$_2$ termination layer.}
    \label{fig:SNO-relaxSr}
\end{figure}

To check the impact of ionic relaxation on the electronic transport we here evaluate the transmission of 2 u.c. SrNbO$_3$ with different vacancies before and after relaxation. The results for O, Nb, and Sr vacancy are shown in Figures \ref{fig:SNO-relaxO}, \ref{fig:SNO-relaxNb}, and \ref{fig:SNO-relaxSr}, respectively. The vacancies are placed in the first unit cell closest to the NbO$_2$ termination (other configurations show comparable sensitivity to relaxation). In the pristine cases, ``relaxed'' indicates ionic relaxation of whole slab with in-plane lattice parameter fixed to bulk value. For defective systems, the ``relaxed'' results are from pristine relaxed slabs with further ionic relaxation after introduction of vacancies. Only ions within a radius of $1.2a$ from the vacancy were further relaxed. While the relaxations typically led to bond angles deviating ca. \SI{5}{\degree} (from the cubic value of \SI{180}{\degree}), no significant change was observed in the conductances near the Fermi level. This justified us to neglect relaxation effects in the main text.

\begin{figure}
    \centering
    \includegraphics[width=0.49\textwidth]{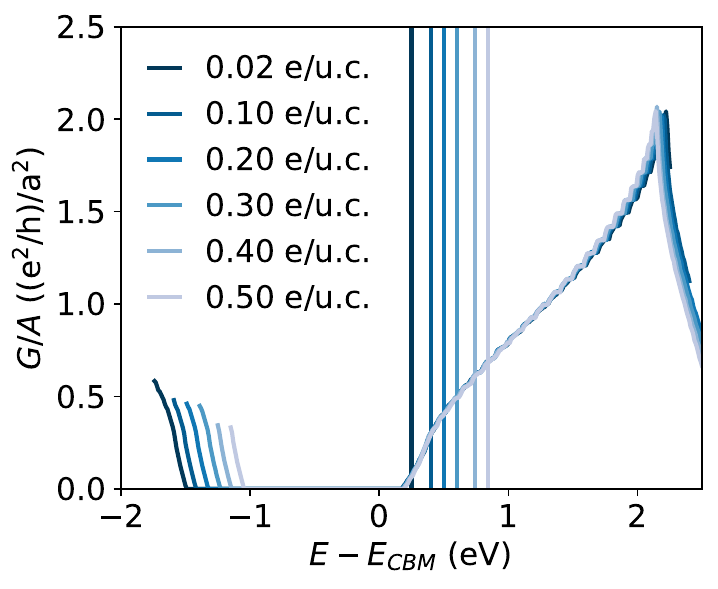}
    \caption{Conductance of SrTiO$_3$ with varying degrees of V doping. The doping is given in number of extra electrons per unit cell. The energies are shifted to conduction band minimum and vertical lines correspond to the different Fermi levels.}
    \label{fig:STO-bulk}
\end{figure}

\section{Effect of doping on conductance in SrTiO$_3$}\label{sec:doping}
To simulate n-type SrTiO$_3$, we dope Ti with V using the VCA. Figure~\ref{fig:STO-bulk} presents the conductances of bulk n-type SrTiO$_3$ with varying V doping, and therefore varying electron count. The conductance due to the conduction band remains almost unaffected by the doping, the major difference is the shift in Fermi energy, as expected.

\begin{figure}
    \centering
    \includegraphics[width=0.5\textwidth]{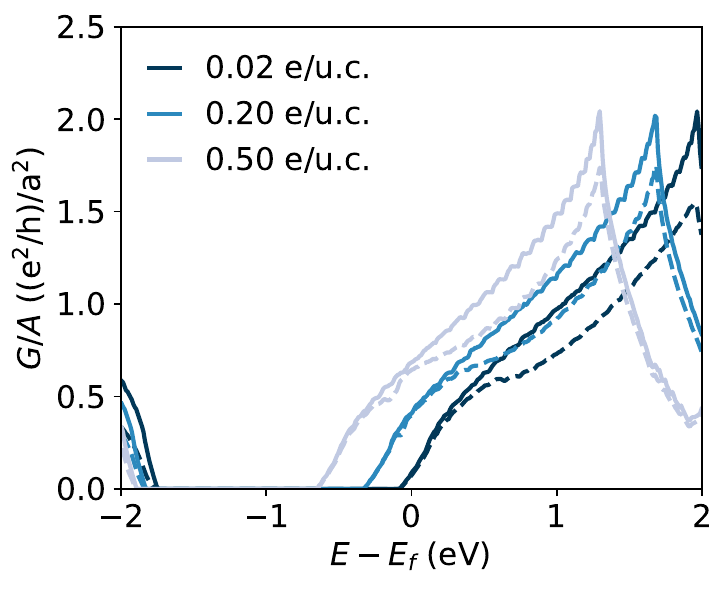}
    \caption{Conductance of SrTiO$_3$ with O vacancy in TiO$_2$ layer (dashed lines). The solid line corresponds to pristine system. Three doping concentrations are shown for simplicity.}
    \label{fig:STO-bulk-O-vac}
\end{figure}

\begin{figure}
    \centering
    \includegraphics[width=0.5\textwidth]{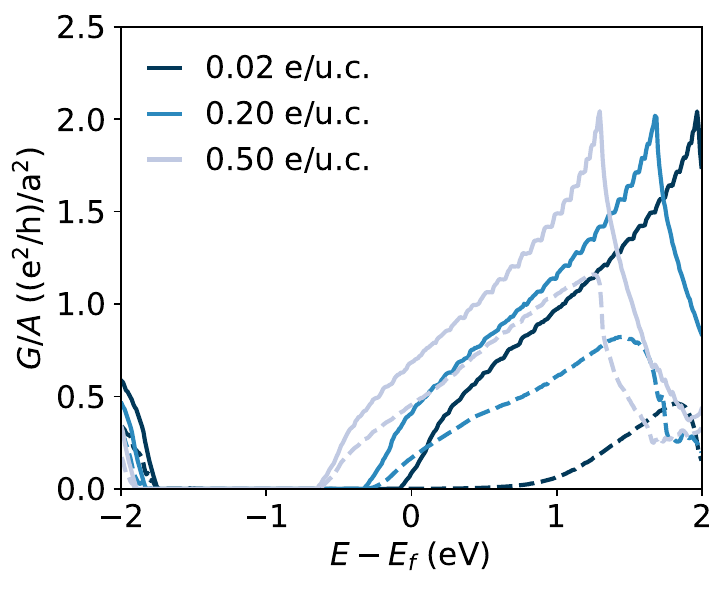}
    \caption{Conductance of SrTiO$_3$ with Ti vacancy (dashed lines). The solid line corresponds to pristine system. Three doping concentrations are shown for simplicity.}
    \label{fig:STO-bulk-TiV-vac}
\end{figure}

\begin{figure}
    \centering
    \includegraphics[width=0.5\textwidth]{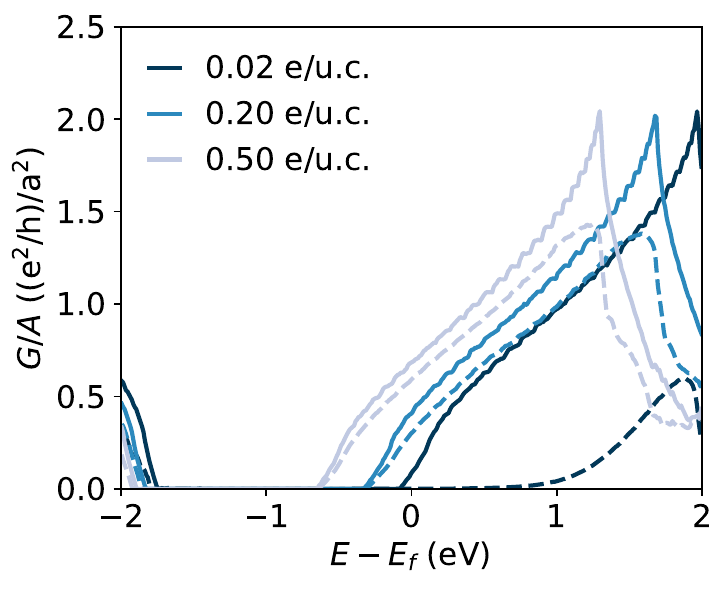}
    \caption{Conductance of SrTiO$_3$ with Sr vacancy (dashed lines). The solid line corresponds to pristine system. Three doping concentrations are shown for simplicity.}
    \label{fig:STO-bulk-Sr-vac}
\end{figure}

Next, the conductance of n-type SrTiO$_3$ with O, Ti, and Sr vacancy is presented in Figs.~\ref{fig:STO-bulk-O-vac}, ~\ref{fig:STO-bulk-TiV-vac}, and ~\ref{fig:STO-bulk-Sr-vac}, respectively. For simplicity, only a few electron concentrations are shown. The energy resolved conductance of n-type SrTiO$_3$ with O vacancy (Fig.~\ref{fig:STO-bulk-O-vac}) is almost unaffected by a change of electron concentration except for a shift in Fermi level. The SrTiO$_3$ with Ti and Sr vacancies, however, have a more complex relation between conductance and doping. The drop in conductance due to one Ti vacancy increases with a decrease in carrier concentration (see Fig.~\ref{fig:STO-bulk-TiV-vac}). This is true not only for the Fermi level, but also for other energies in the conduction band. Similar trends are observed for SrTiO$_3$ with Sr vacancy, however, the drop in conductance is smaller for highly doped systems compared to Ti vacancy, while in lowly doped systems the conductances are comparable. Hence, the Sr vacancy has smaller scattering strength, compared to Ti vacancy, in highly doped systems, while they are comparable in lowly doped systems (as seen in main text Figure~\ref{fig:STO-thickness-conc-scatter}).

\clearpage
\bibliography{references}

\end{document}